\documentclass[prd, nofootinbib, preprint, 12 pt]{revtex4}
\usepackage{amssymb}
\usepackage{amsmath,graphicx,color,epsfig}
\usepackage{pstricks}
\usepackage{float}
\usepackage{subfigure}
\begin{document}
\title{Form factors $f^{B\to \pi}_+(0)$ and
$f^{D\to \pi}_+(0)$ in $QCD$ and determination of $|V_{ub}|$ and
$|V_{cd}|$}
\author{Zuo-Hong Li$~^{a,}$\footnote{lizh@ytu.edu.cn}, Nan Zhu~$^{a}$, Xiao-Jiao Fan~$^{a}$ and Tao
Huang~$^{a,~b,~}$\footnote{huangtao@ihep.ac.cn}}\affiliation{{\small\it $^a$Department of Physics, Yantai University, Yantai 264005, P.R.China\\
$^b$Institute of High Energy Physics and Theoretical Physics Enter
for Science Facilities, Chinese Academy of Sciences, Beijing
100049,P.R.China}}
\date{\today}
\begin{abstract}
We present a QCD study on $B, D\to\pi$ semileptonic transitions at
zero momentum transfer and an estimate of magnitudes of the
associated CKM matrix elements. Light cone sum rules (LCSRs) with
chiral correlator are applied to calculate the form factors $f^{B\to
\pi}_+(0)$ and $f^{D\to \pi}_+(0)$. We show that there is no twist-3
and-5 component involved in the light-cone expansions such that the
resulting sum rules have a good convergence and offer an
understanding of these form factors at twist-5 level. A detailed
$\mathcal{O}(\alpha_s)$ computation is carried out in leading
twist-2 approximation and the $\overline{MS}$ masses are employed
for the underlying heavy quarks. With the updated inputs and
experimental data, we have $f^{B\to \pi}_+(0)=0.28^{+0.05}_{-0.02}$
and $|V_{ub}|=(3.4^{+0.2}_{-0.6}\pm0.1\pm0.1)\times10^{-3}$;
$f^{D\to \pi}_+(0)=0.62\pm0.03$ and
$|V_{cd}|=0.244\pm0.005\pm0.003\pm0.008$. As a by-product, a
numerical estimate for the decay constant $f_D$ is
yielded as $f_{D}=190^{+12}_{-11}\mathrm{MeV}$.
\end{abstract}
\maketitle
\section{Introduction}
An intensive study on the Cabibbo-Kobayashi-Maskawa (CKM) matrix
elements remains a cornerstone of high energy physics programme, in
testing the standard model (SM) and exploring new physics. The
unitarity of the CKM matrix must be put to a test by
phenomenological research on the so called unitarity triangle.
Opposite to the side of the triangle whose length depends on, in
addition to the CKM parameters $|V_{cb}|$ and $|V_{ud}|$, the
elements involving heavy-light quark mixing $|V_{ub}|$ and
$|V_{cd}|$, the angle $\beta$ is presently well-measured. So
precision determination of them is central to the unitarity testing.
Exclusive processes offer an indispensable avenue to understand
these parameters. The decays of heavy mesons into a light
pseudoscalar meson plus an electron and its antineutrino can proceed
at electro-weak tree level and are much less sensitive to new
physics, and accordingly they could serve as preferred exclusive
channels to probe both elements that we take interest in, namely,
$|V_{ub}|$ and $|V_{cd}|$. Then we are confronted with calculation
of the hadronic matrix elements, say, that for the
$\overline{B^0}\to\pi^+$ transition parameterized usually as
\begin{eqnarray}
\langle \pi(p)|\bar{u}\gamma_{\mu}b|B(p+q)\rangle=2f^{B\to
\pi}_+(q^{2})p_{\mu}+(f^{B\to \pi}_+(q^{2})+f^{B\to
\pi}_-(q^{2}))q_{\mu},
\end{eqnarray}
with the momentum assignment specified in brackets, and $f^{B\to
\pi}_{+}(q^{2})$ and $f^{B\to \pi}_{-}(q^{2})$ being the form
factors describing QCD dynamics in the decay, of which only the
former is related if the small electron mass is neglected. Combining
the partial rates measured in some $q^2$ bins with the form factor
predictions of different QCD approaches, one could achieve the
values for related $|V_{ij}|$. Another approach is to fit the
experimental observations using the various form factor
parameterizations. In such way, a strong constraint is imposed on
$q^2$ distributions of the form factors such that one may obtain a
precise estimate of the products $f_+ (0)|V_{ij}|$, in which case
theoretical task boils down to estimating the form factors $f_+ (0)$
at $q^2=0$. Requiring a good knowledge of the form factors, the
exclusive avenues to $|V_{ij}|$ are theoretically more challenging
than inclusive approaches. The continual data updates have aroused
one's enthusiasm for exploring heavy-to-light transitions to
approach an understanding of the CKM parameters. In the wake of the
recent accurate measurements of the semileptonic processes by the
BaBar \cite{P.del 052011,P.del 032007} and CLEO \cite{D.Besson,
B.I.Eisenstein} collaborations, new progress has been achieved in
this respect. Some extent of tension, however, still holds between
inclusive and exclusive extractions of $|V_{ub}|$. A global
data-fitting from CKMFitter \cite{ckmf} and UTfit \cite{utf} is in
favor of a smaller $|V_{ub}|$ than inclusive determinations. One can
be referred to \cite{CKM} for a comprehensive overview of the
current status of the CKM matrix elements.

Developed from QCD sum rule technique, light cone sum rules
(LCSRs)\cite{LCSR1, LCSR2} have become a powerful competitor in
making predictions for heavy to light transitions. Complementary to
lattice QCD (LQCD) simulations, this approach is successfully
applied to study $B$ decays\cite{LCSR2, LCSR, pball05, pballplb,
G.Duplancic08, A.Khodjamirian11, Huang1, Huang2}: whereas the former
are available for the high $q^2$, LCSR calculation is applicable for
the low and intermediate $q^2$. Utilizing the LCSR predictions for
$f^{B\to \pi}_+(q^2)$, one has launched a painstaking investigation
into $|V_{ub}|$ \cite{pball05, pballplb, G.Duplancic08,
A.Khodjamirian11}, with a consistent result with those using LQCD.
The same approach has also been taken to understand $D\to \pi,K$
decays in \cite{P.Ball06, A.Khodjamirian09}, the resulting sum rules
\cite{A.Khodjamirian09} being employed to extract $|V_{cd}|$ and
$|V_{cs}|$ .

The uncertainties in the light meson distribution amplitudes (DAs)
involved in the sum rules, however, would have different degrees of
impacts on the results. To gain enlightenment on how to further
improve accuracy of the LCSR calculations, it is essential to look
into the role played by each of the higher twist DAs. A systematic
numerical analysis shows that whereas the twist-4 effects account
for only a few percent of the total sum rule results, the chirally
enhanced twist-3 contributions are numerically large enough to be
comparable with the twist-2 ones in the $B$ meson cases, and even
about twice as large as the latter for $D$ decays. As a result,
there are a few problems left unsolved. To start with, one might
doubt whether the potential twist-5 effects are negligible in
particular while assessing $D$ decays. Secondly, the sum rule
pollution by twist-3 would be serious on account of the combined
uncertainties of the DAs and chiral enhancement factor. Finally,
since there is an extremely different sensitivity to twist-2 between
the sum rules for $f^{B\to \pi}_+(0)$ and $f^{D\to \pi}_+(0)$, a
successful LCSR application to the latter does not necessary assure,
with the same inputs, a reliable LCSR prediction for the former. For
the moment, these issues are difficult to essentially settle within
the LCSR framework. The trick suggested in \cite{LCSR2,Huang1} is
available as a temporary scenario to approach them.

Focusing on $f^{B\to \pi}_+(0)$ and $f^{D\to \pi}_+(0)$, in this
work we intend to reconsider heavy to light transitions in the
revised LCSR version so as to provide a calculation independent of
the traditional LCSR ones and further a determination of the
associated CKM parameters. We will expound that this approach does
not involve the twist-3 and-5 DAs, which enables us to get an
understanding of the form factors to twist-5 precision only
resorting to the known twist-2 and -4 DAs and to perform a
cross-check between the resulting LCSRs for $f^{B\to \pi}_+(0)$ and
$f^{D\to \pi}_+(0)$. This paper is organized as follows. In the
following Section we put forward our derivation of the sum rules in
question, including a detailed next-to-leading order (NLO) QCD
calculation in twist-2 approximation, and elaborate on the key
technical points. The modifications and improvements made are also
addressed in comparison with the previous calculations
\cite{Huang1,Huang2}. In Section 3, after discussing assignment of
the parameters for which updated and consistent findings are
selected as inputs, we shift into numerical computation with a
systematic error discussion included, by means of up-to-date
experimental data, and present our LCSR results for $f^{B\to \pi}_+
(0)$ and $f^{D\to \pi}_+ (0)$ and the determination of $|V_{ub}|$
and $|V_{cd}|$. Too we report on an estimate of the decay constant
$f_D$, as a by-product. The final Section is devoted to a summary.
\section{ QCD calculation of  $f^{B\to \pi}_+ (0)$ and $f^{D\to \pi}_+ (0)$}
The starting point of LCSR calculation is to consider a correlation
function with $T$ product of currents sandwiched between the vacuum
and a light meson state $L$. In the coordinate space and for large
and negative virtuality of the current operators, the correlation
function can be in form expanded, in the small light cone distance
$x^2\approx0$, as,
\begin{eqnarray}
 correlation~function\sim\sum_mC_m(x)\langle
 L(p)|\mathcal{O}_m(x,0)|0\rangle,
\end{eqnarray}
where $C_m(x)$ are the Wilson coefficients, $\mathcal{O}_m(x,0)$ the
nonlocal operators built out of quark and/or gluon fields, and the
matrix elements $\langle L(p)|\mathcal{O}_m(x,0)|0\rangle$ have an
expansion form in term of the light cone DAs $\Psi^{(n)}$ with
increasing twist $n$. The power series $\sum C_n(x\cdot p)^n$
($x\cdot p \sim 1 $ for a large external momentum $p$) appearing in
the expansion process are summed up effectively, which works out
some of the problems with the expansion in the small distance
$x\approx 0$. Switching (2) to momentum space, we have
\begin{eqnarray}
 correlation~function\sim\sum_nT^{(n)}_H\otimes\Psi^{(n)},
\end{eqnarray}
a factorized form with the hard kernel $T^{(n)}_H$ being convoluted
with $\Psi^{(n)}$. Whereas the process-independent $\Psi^{(n)}$
parameterize the long distance effects below a factorization scale
$\mu$, the process-dependent amplitudes $T^{(n)}_H$ describe the
hard-scattering dynamics above $\mu$, which are perturbatively
calculable and have the following expansions in $\alpha_s$:
\begin{eqnarray}
 T^{(n)}_H=
 T^{(n)}_0+\frac{\alpha_sC_F}{4\pi}T^{(n)}_1+\cdot\cdot\cdot.
\end{eqnarray}
If calculation is restricted to $O(\alpha_s)$ accuracy, we need just
to estimate the leading order (LO) contributions $T^{(n)}_o$ and NLO
corrections $T^{(n)}_1$. Then the remaining procedure is standard.

Now let us take up our LCSR calculations of $f^{B\to \pi}_+(0)$ and
$f^{D\to \pi}_+(0)$. Allowing for the similarity of the two
situations, for definiteness we would like to concentrate on the
former. Moreover, throughout the paper the chiral limit $m_{\pi}=0$
is taken. We follow \cite{LCSR2,Huang1} and adopt the following
correlation function:
\begin{eqnarray}
\Pi_{\mu}(p,q)&=&i\int d^{4}x e^{iqx}\langle
\pi(p)|T\{J_{\mu}^{V+A}(x), J_B^{P+S}(0)\}|0\rangle\nonumber\\
&=&F((p+q)^{2})p_{\mu}+\tilde{F}((p+q)^{2})q_{\mu}.
\end{eqnarray}
Here we substitute the chiral currents
$J_{\mu}^{V+A}(x)=\bar{u}(x)\gamma_{\mu}(1+\gamma_{5})b(x)$ and
$J_B^{P+S} =m_{b}\bar{b}(0)i(1+\gamma_{5})d(0)$, respectively, for
the operators adopted usually
$J_{\mu}(x)=\bar{u}(x)\gamma_{\mu}b(x)$ and $J_B
=m_{b}\bar{b}(0)i\gamma_{5}d(0)$. The operator replacements do not
violate renormalization group invariance of the correlation
function, for both $J_{\mu}^{V+A}$ and $J_B^{P+S}$, like the latter
two, have an anomalous dimension of zero, and however make the
correlation function receive an additional contribution from the set
of scalar $B$ mesons. In view of that the mass of the lowest scalar
$B$ meson is slightly below the one of the first excited state of
the pseudoscalar $B$ mesons, we could safely isolate the pole term
of the pseudoscalar ground state from the contributions of higher
resonances and continuum states.

For the present purpose, it is sufficient to consider the part
proportional to $p_{\mu}$ in (5), that is, the invariant function
$F((p+q)^{2})$. It has the pole term of interest to us,
\begin{eqnarray}
F_{pole}((p+q)^{2})=\frac{2m_{B}^{2}f^{B\to
\pi}_+(0)f_{B}}{m_{B}^{2}-(p+q)^{2}},
\end{eqnarray}
where $m_B$ and $f_B$ indicate, respectively, the $B$ meson mass and
decay constant defined as
\begin{eqnarray}
\langle B|\bar{b}i\gamma_{5}d|0\rangle=\frac{m_{B}^{2}f_{B}}{m_{b}}.
\end{eqnarray}
The spectral function $\rho^H(s)$ is introduced to reckon in the
higher state contributions in a dispersion integral starting with
the threshold $s_0^B$, which should be assigned near the squared
mass of the lowest scalar $B$ meson. At this point, what remains to
be done is the light cone expansion calculation on $F((p+q)^{2})$,
from which the corresponding QCD spectral function $\rho^{QCD}(s)$
is extracted in order to get the sum rule for $f^{B\to \pi}_+(0)$ by
matching the Borel improved theoretical and phenomenological forms
with the duality assumption
$\rho^H(s)=\rho^{QCD}(s)\Theta(s-s_0^B)$.
\begin{figure}
\includegraphics[scale=1.1]{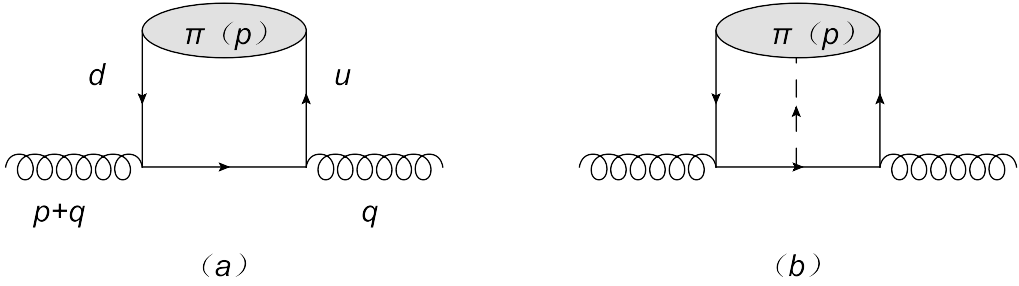}
\begin{center}
\caption{Tree-level Feynman diagrams contributing to the correlation
function.}
\end{center}
\end{figure}

The light cone expansion of (5) goes effectively in the large
space-like momentum region $(p+q)^{2}-m_{b}^{2}<<0$ for the
$b\bar{d}$ channel. At tree-level and to NLO in the light-cone
expansion of the $b$ quark propagator, it can be illustrated by the
two Feynman diagrams as depicted in Fig.1. In comparison with
Fig.1(a), which corresponds to the leading term in the quark
propagator and illustrates the two-particle contribution, Fig.1(b)
portrays the three-particle Fock state effect due to the
soft-emission correction to the free quark propagator, which is
expressed as
\begin{eqnarray}
-i g_{s}\int\frac{d^{4}k}{(2\pi)^{4}} e^{-ikx}
\int^{1}_{0}dv\Big[\frac{1}{2}\frac{\not{k}+m_{b}}{(m_{b}^{2}-k^{2})^{2}}G^{\mu\nu}(vx)\sigma_{\mu\nu}
+\frac{1}{m_{b}^{2}-k^{2}}v x_{\mu}G^{\mu\nu}(vx)\gamma_{\nu}\Big].
\end{eqnarray}
The contribution of Fig.1(a) to $F((p+q)^{2})$ is easy to estimate,
using the definition of the pionic two-particle DAs:
\begin{eqnarray}
\langle\pi(p)|\bar{u}_{\alpha}(x)d_{\beta}(0)|0\rangle_{x^2\to
0}&=&i\frac{f_{\pi}}{4}\int_0^1du~e^{iup\cdot
x}\Big[(/\kern-0.57em p\gamma_5)_{\beta\alpha}\varphi_{\pi}(u)\nonumber\\
&-&(\gamma_5)_{\beta\alpha}\mu_{\pi}\phi^p_{3\pi}(u)+\frac{1}{6}(\sigma_{\xi\eta}\gamma_5)_{\beta\alpha}p^{\xi}x^{\eta}\mu_{\pi}\phi^{\sigma}_{3\pi}(u)\nonumber\\
&+&\frac{1}{16}(/\kern-0.57em
p\gamma_5)_{\beta\alpha}x^2\phi_{4\pi}(u)-i\frac{1}{2}(/\kern-0.57em
x\gamma_5)_{\beta\alpha}\int_0^u\psi_{4\pi}(v)dv\Big],
\end{eqnarray}
where $u$ is the fraction of the light cone momentum $p_0+p_3$ of
the pion carried by the constituent $u$ quark. While
$\varphi_{\pi}(u)$ denotes the twist-2 DA, both $\phi^p_{3\pi}(u)$
and $\phi^{\sigma}_{3\pi}(u)$, which are accompanied by the chiral
enhancement factor $\mu_{\pi}$, have twist-3, and the other two
functions are both of twist-4. From the following trace form, which
emerges obviously as one works in the momentum space,
\begin{eqnarray}
Tr\{\underbrace{[d(\bar{u}p)\bar{u}(up)]}_{wavefunction}\gamma_{\mu}(1+\gamma_5)(/\kern-0.57em
q+u/\kern-0.57em p+m_b)(1+\gamma_5)\},
\end{eqnarray}
we see readily that the twist-3 components make a vanishing
contribution to the light cone expansion, because of the
corresponding Dirac wavefunctions. In fact, the same happens to the
three-particle situation, as shown from a straightforward
computation with (8) and the decomposition:
\begin{eqnarray}
&&\langle\pi(p)|\bar{u}_{\alpha}(x)g_sG_{\mu\nu}(vx)d_{\beta}(0)|0\rangle_{x^2\to
0}=\frac{1}{4}\int\mathcal{D}\alpha_i e^{ip\cdot x(\alpha_1+\alpha_3v)}\Big[if_{3\pi}(\sigma^{\rho\lambda}\gamma_5)_{\beta\alpha}\nonumber\\
&&\times(p_{\mu}p_{\rho}g_{\nu\lambda}-p_{\nu}p_{\rho}g_{\mu\lambda})\Phi_{3\pi}(\alpha_i)-f_{\pi}(\gamma^{\rho}\gamma_5)_{\beta\alpha}\Big\{(p_{\nu}g_{\mu\rho}-p_{\mu}g_{\nu\rho})\Psi_{4\pi}(\alpha_i)\nonumber\\
&&+\frac{p_{\rho}(p_{\mu}x_{\nu}-p_{\nu}x_{\mu})}{p\cdot
x}(\Phi_{4\pi}(\alpha_i)+\Psi_{4\pi}(\alpha_i))\Big\}-i\frac{f_{\pi}}{2}\epsilon_{\mu\nu\delta\lambda}(\gamma_{\rho})_{\beta\alpha}\nonumber\\
&&\times
\Big\{(p^{\lambda}g^{\delta\rho}-p^{\delta}g^{\lambda\rho})\widetilde{\Psi}_{4\pi}(\alpha_i)+\frac{p^{\rho}(p^{\delta}x^{\lambda}-p^{\lambda}x^{\delta})}
 {p\cdot
 x}\Big(\widetilde{\Phi}_{4\pi}(\alpha_i)+\widetilde{\Psi}_{4\pi}(\alpha_i)\Big)\Big\}\Big],
\end{eqnarray}
where $G_{\mu\nu}$ is the gluonic field strength tensor and
$\mathcal{D}\alpha_i=d\alpha_1d\alpha_2d\alpha_3\delta(1-\alpha_1-\alpha_2-\alpha_3)$;
$\Phi_{3\pi}(\alpha_i)$ indicates the twist-3 component of the
three-particle DAs, and the remaining functions are all of twist-4.
In the usual LCSR application to $D$ decays, the chirally enhanced
twist-3 terms provide a leading contribution, which engenders much
negative influence as aforementioned.

At present, the two-particle contribution $F_0^{(2p)}((p+q)^2)$ can
be written down in a form that the DAs are convoluted with the
corresponding LO hard scattering amplitudes,
\begin{eqnarray}
F_{0}^{(2p)}((p+q)^2)&=&-f_{\pi}\int_0^1du~\big[T^{(2)}_{0}((p+q)^2,u)\varphi_{\pi}(u)\nonumber\\
&-&T^{(4)}_0((p+q)^2,u)\int_0^u\psi_{4\pi}(v)dv-\widetilde{T}^{(4)}_0((p+q)^2,u)\phi_{4\pi}(u)\big],
\end{eqnarray}
with
\begin{eqnarray}
&&T^{(2)}_{0}((p+q)^2,u)=-2\frac{m_b^2}{m_{b}^{2}-u(p+q)^{2}},\\
&&T^{(4)}_{0}((p+q)^2,u)=2\frac{u}{(m_b^2-u(p+q)^2)}\Big(u\frac{d}{du}+1\Big),\\
&&\widetilde{T}^{(4)}_{0}((p+q)^2,u)=-\frac{u^2}{2(m_b^2-u(p+q)^2)}\frac{d^2}{du^2}.
\end{eqnarray}
The three-particle contribution is of the following convolution
\begin{eqnarray}
F_{0}^{(3p)}((p+q)^2)=-f_{\pi}\int_0^1du~\overline{T}^{(4)}_0((p+q)^2,u)I_{4\pi}(u),
\end{eqnarray}
with
\begin{eqnarray}
\overline{T}^{(4)}_0((p+q)^2,u)=2\frac{u}{(m_b^2-u(p+q)^2)}\frac{d}{du},\\
\end{eqnarray}
and
\begin{eqnarray}
I_{4\pi}(u)&=&\int_0^u~d\alpha_1~~\int_{(u-\alpha_1)/(1-\alpha_1)}^1
\frac{dv}{v}\left[2\Psi_{4\pi}(\alpha_i)+2\widetilde{\Psi}_{4\pi}(\alpha_i)\right.\nonumber\\
&-&\left.\left.\Phi_{4\pi}(\alpha_i)-\widetilde{\Phi}_{4\pi}(\alpha_i)\right]\right|_{\begin{subarray}{l}
 \alpha_{2}=1-\alpha_{1}-\alpha_{3}\\
 \alpha_{3}=(u-\alpha_{1})/\upsilon\end{subarray}}.
\end{eqnarray}
Then we can attain the imaginary part of $F_0^{QCD}((p+q)^2)=
F_{0}^{(2p)}((p+q)^2)+F_{0}^{(3p)}((p+q)^2)$ via estimating the ones
of the hard kernels in (13--15) and (17), and further the desired
QCD spectral function $\rho^{QCD}_0(s)$.  The result is as follows:
\begin{eqnarray}
 \rho^{QCD}_0(s)&=&2f_{\pi}\int_0^1du\delta\big(1-u\frac{s}{m_b^2}\big)\left[\varphi_{\pi}(u)+\frac{u}{m_b^2}\Big(u\frac{d}{du}+1\Big)\int_0^u\psi_{4\pi}(v)dv\right.\nonumber\\
&-&\left.\frac{u^2}{4m_b^2}\frac{d^{2}}{du^2}\phi_{4\pi}(u)-\frac{u}{m_b^2}\frac{d}{du}I_{4\pi}(u)\right].
\end{eqnarray}

At twist-4 level, we have provided a complete LO light cone QCD
representation for $F((p+q)^2)$. In its present form, the ensuing
continuum substraction could be enforced systematically for the
twist-4 as well as twist-2 parts, with the known QCD spectral
function. This improves explicitly the previous treatment
\cite{Huang1, Huang2} in which the twist-4 contribution is written
down in a form not suitable for continuum substraction.
\begin{figure}
\includegraphics[scale=1.1]{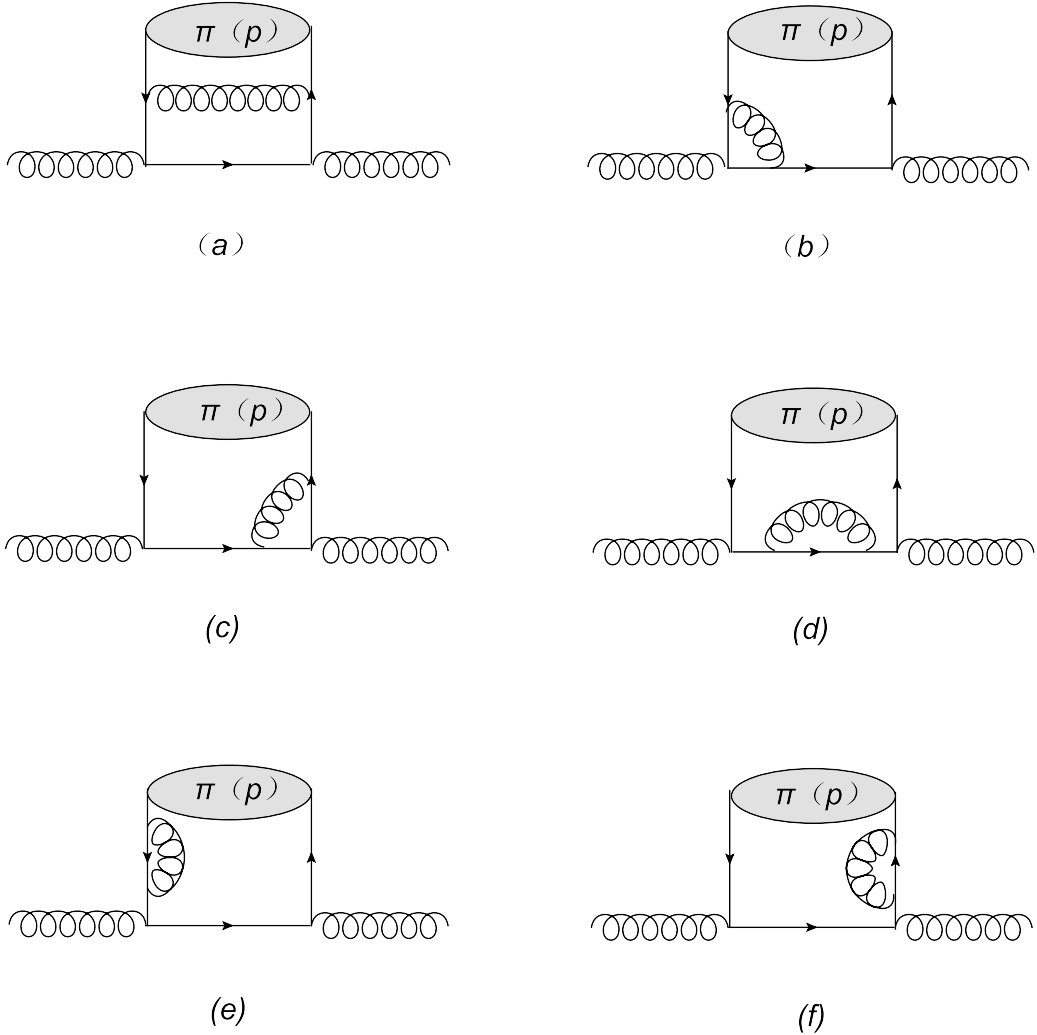}
\caption{One-loop Feynman diagrams contributing to the correction
function.}
\end{figure}

Our main task is to evaluate the gluon emission effect on the LCSR
for $f_+^{B\to \pi}(0)$ at one loop level. It should be sufficient
for this purpose to calculate the NLO parts of the leading twist-2
and the chirally enhanced twist-3 contributions. To be specific, we
are about to compute the six Feynman diagrams plotted in Fig.2, to
the accuracy in question. Fig.2(a) depicts diagrammatically the
hard-exchange correction between the outgoing and spectator quarks
in the $B\to \pi$ transition. From the nature of the correlation
function, we deduce easily that there is no UV divergence in
Fig.2(a) or it could not be canceled out. Of the other figures,
Figs.2(b,~e) and Figs.2(c,~f) involve, respectively, the partial
one-loop contributions to the $1+\gamma_5$ vertex and to the
$\gamma_{\mu}(1+\gamma_5)$ one, while Fig.2(d) does the remaining
loop contribution to both operators. It is conceivable that each of
these five includes both UV and IR divergences, except Fig.2(d)
which is merely UV divergent because obviously if any IR divergence
arises it can not be reasonably absorbed into a pionic DA.

It is found that the twist-3 components still produce no effect at
one-loop level, for the same reason as in the tree-level case. Hence
the NLO computation is reduced to a calculation of the
$\mathcal{O}(\alpha_s)$ correction to the LO twist-2 contribution
$T^{(2)}_{0}$ (for brevity, hereafter we indicate the LO twist-2
contribution by the symbol $T_0$ instead of $T^{(2)}_{0}$ and the
corresponding NLO correction by $T_1$, up to a prefactor
$\alpha_s/4\pi$). We work in the Feynman gauge. In addition, we use
the dimensional regularization and $\overline{MS}$ scheme to deal
with the ultraviolet (UV) and infrared (IR) divergences appearing in
the calculation, such that the LO evolution kernel of
$\varphi_{\pi}(u)$\cite{G.P.Lepage} achieved early in the same
prescription is available for a proof of QCD factorization for the
resulting twist-2 contribution to $F((p+q)^2)$ as we attempt to
segregate the long distance contribution from the perturbative
kernel. The calculation is tedious and complicated. Here we present,
for the first time, some details of the diagram calculation. We
summarize the divergence contribution to $T_1$ from each of the
diagrams in Fig.2 as follows,
\begin{eqnarray}
&&T_{1(a)}^{\mathrm{div}}(u,r)=4\frac{1}{\overline{u}r^{2}}\left[(1-r)\mathrm{ln}(1-r)-\left(\frac{1}{u}-r\right)\mathrm{ln}(1-ur)\right]\Delta_{\mathrm{IR}},\\
\nonumber\\
&&T_{1(b)}^{\mathrm{div}}(u,r)=4\frac{1}{1-ur}\left[\left(\frac{r-1}{\overline{u}r}\mathrm{ln}\frac{1-ur}{1-r}+1\right)\Delta_{\mathrm{IR}}-2\Delta_{\mathrm{UV}}\right],\\
\nonumber\\
&&T_{1(c)}^{\mathrm{div}}(u,r)=4\frac{1}{1-ur}\left[\left(\frac{1}{ur}\mathrm{ln}(1-ur)+1\right)\Delta_{\mathrm{IR}}-\frac{1}{2}\Delta_{\mathrm{UV}}\right],\\
\nonumber\\
&&T_{1(d)}^{\mathrm{div}}(u,r)=4\frac{2+ur}{(1-ur)^{2}}\Delta_{\mathrm{UV}},\\
&&T_{1(e+f)}^{\mathrm{div}}(u,r)=-\frac{2}{1-ur}\Delta_{\mathrm{IR}}+\frac{2}{1-ur}\Delta_{\mathrm{UV}}.
\end {eqnarray}
Here $\bar u=1-u$, $r=(p+q)^{2}/m_{b}^{2}$ and
\begin{eqnarray}
\Delta_{\mathrm{IR}}(\Delta_{\mathrm{UV}})=\frac{1}{\varepsilon_{\mathrm{IR}}}\left(\frac{1}{\varepsilon_{\mathrm{UV}}}\right)-\gamma_{E}+\mathrm{ln}4\pi
\end{eqnarray}
with the $\varepsilon_{\mathrm{UV}}$ and $\varepsilon_{\mathrm{IR}}$
introduced to regularize the UV and IR divergences, respectively.
Obviously, the yielded results are as expected.

Adding all the divergent and finite terms together, we have the NLO
correction
\begin{eqnarray}
T_1(u,r)&=&2\bigg\{\frac{1}{1-ur}\left(3-2~\mathrm{ln}(1-r)\frac{1-r-ur}{ur^{2}}-2~\mathrm{ln}\left(\frac{1-ur}{1-r}\right)\frac{1-r-u\overline{u}r^{2}}{u\overline{u}r^{2}}\right)\Delta_{\mathrm{IR}}\nonumber\\
&+&\frac{6ur}{(1-ur)^{2}}\Delta_{\mathrm{UV}}+\frac{1+ur}{(1-ur)^{2}}\left(3-3~\mathrm{ln}\frac{m_{b}^{2}}{\mu^{2}}+\frac{1}{ur}\right)\nonumber\\
&+&2\left[\frac{1}{\overline{u}r}-\frac{1}{r(1-ur)}-\left(\frac{1}{\overline{u}r^{2}}-\frac{1}{1-ur}\right)\mathrm{ln}\frac{m_{b}^{2}}{\mu^{2}}\right]\mathrm{ln}(1-r)\nonumber\\
&+&2\left(\frac{1}{1-ur}-\frac{1}{\overline{u}r^{2}}\right)\left(\mathrm{ln}^{2}(1-r)+\mathrm{Li}_{2}(r)\right)\nonumber\\
&+&\left[\frac{4}{1-ur}+\frac{ur+u^{2}r+\overline{u}}{\overline{u}(ur)^{2}}-2\left(\frac{2}{1-ur}-\frac{1-\overline{u}r}{u\overline{u}r^{2}}\right)
\mathrm{ln}\frac{m_{b}^{2}}{\mu^{2}}\right]\mathrm{ln}(1-ur)\nonumber\\
&-&2\left(\frac{2}{1-ur}-\frac{1-\overline{u}r}{u\overline{u}r^{2}}\right)\left(\mathrm{ln}^{2}(1-ur)+\mathrm{Li}_{2}(ur)\right)\bigg\},
\end {eqnarray}
with the dilogarithm
$\mathrm{Li}_2(x)=-\int_0^x~dt~\frac{\mathrm{ln}(1-t)}{t}$.

Keep in mind that up to now the quark mass has been treated as a
bare quantity. A mass renormalization must be performed in the
$\overline{MS}$ scheme, in order to have a UV renormalized
hard-scattering amplitude $T$ via adding $T_1$ to $T_0$. It can be
done by making the parameter replacement $m_b\to Z_m m_b$ in the
related expressions, with the renormalization constant
$Z_m=1-3\Delta_{\mathrm{UV}}\frac{\alpha_sC_F}{4\pi}$. As a result,
the tree level expression (13), to the accuracy required, is
modified to the form
\begin{eqnarray}
T_{0}(u,r)=\frac{2}{ur-1}-\frac{\alpha_{s}C_{F}}{4\pi}\frac{12ur}{(1-ur)^{2}}\Delta_{\mathrm{UV}},
\end{eqnarray}
but the NLO term $T_{1}(u,r)$ keeps its form unchanged. Here $m_b$
entering $r$ should be understood as the $\overline{MS}$ mass. The
additional UV divergent contribution in (28), as it should be,
precisely cancels out the one of (27). Then a complete UV
renormalized result is obtained as
\begin{eqnarray}
T(u,r)&=&T_{0}(u,r)+\frac{\alpha_{s}C_{F}}{4\pi}T_{1}(u,r)\nonumber\\
&=&2\bigg\{\frac{1}{1-ur}\left(3-2~\mathrm{ln}(1-r)\frac{1-r-ur}{ur^{2}}-2~\mathrm{ln}\left(\frac{1-ur}{1-r}\right)\frac{1-r-u\overline{u}r^{2}}{u\overline{u}r^{2}}\right)\Delta_{\mathrm{IR}}\nonumber\\
&+&\frac{1+ur}{(1-ur)^{2}}\left(3-3~\mathrm{ln}\frac{m_{b}^{2}}{\mu^{2}}+\frac{1}{ur}\right)\nonumber\\
&+&2\left[\frac{1}{\overline{u}r}-\frac{1}{r(1-ur)}-\left(\frac{1}{\overline{u}r^{2}}-\frac{1}{1-ur}\right)\mathrm{ln}\frac{m_{b}^{2}}{\mu^{2}}\right]\mathrm{ln}(1-r)\nonumber\\
&+&2\left(\frac{1}{1-ur}-\frac{1}{\overline{u}r^{2}}\right)\left(\mathrm{ln}^{2}(1-r)+\mathrm{Li}_{2}(r)\right)\nonumber\\
&+&\left[\frac{4}{1-ur}+\frac{ur+u^{2}r+\overline{u}}{\overline{u}(ur)^{2}}-2\left(\frac{2}{1-ur}-\frac{1-\overline{u}r}{u\overline{u}r^{2}}\right)
\mathrm{ln}\frac{m_{b}^{2}}{\mu^{2}}\right]\mathrm{ln}(1-ur)\nonumber\\
&-&2\left(\frac{2}{1-ur}-\frac{1-\overline{u}r}{u\overline{u}r^{2}}\right)\left(\mathrm{ln}^{2}(1-ur)+\mathrm{Li}_{2}(ur)\right)\bigg\}.
\end{eqnarray}
We need to add that superior to use of the pole mass for the b quark
\cite{Huang1,Huang2}, employing the $\overline{MS}$ mass could
render not only the calculation free from some element of
uncertainty but the physical meaning more obvious even when the
calculation is performed at QCD tree level, as shown in (28).

To proceed, we embark on handling the IR divergence term,
\begin{eqnarray}
T^{\mathrm{IR}}(u,r)=\frac{2\Delta_{\mathrm{IR}}}{1-ur}\Big[3-2\mathrm{ln}(1-r)\frac{1-r-ur}{ur^2}\nonumber
-2\mathrm{ln}(\frac{1-ur}{1-r})\frac{1-r-u\bar{u}r^{2}}{u\bar{u}r^{2}}\Big].
\end{eqnarray}
If we try to subtract the divergent part from the UV renormalized
hard amplitude to represent the invariant function $F((p+q)^2)$ in
the form of QCD factorization, it has to abide by the form
\begin{eqnarray}
T^{\mathrm{IR}}(u,r)=-\Delta_{\mathrm{IR}}\int_0^1dv~V_0(v,u)~T_0(v,r),
\end{eqnarray}
where $V_0(v,u)$ is the kernel of the evolution equation of the
pionic twist-2 DA \cite{G.P.Lepage}. As checked readily, this is
indeed the case. We can therefore eliminate the divergence by
defining a scale dependent DA as
\begin{eqnarray}
\varphi_{\pi}(u,\mu)=\varphi_{\pi}(u)-\Delta_{\mathrm{IR}}\frac{\alpha_sC_F}{4\pi}\int_0^1dv~V_0(u,v)~\varphi_{\pi}(v),
\end{eqnarray}
which is convoluted with the perturbative kernel
$T^H(u,r,\mu)=T(u,r)|_{\Delta_{\mathrm{IR}}=0}$. As a result, the
twist-2 contribution to $F((p+q)^2)$ observes, at NLO, the following
QCD factorization:
\begin{eqnarray}
 F^{QCD}((p+q)^2)=
 -f_{\pi}\int_0^1du~T^H(u,r,\mu)~\varphi_{\pi}(u,\mu).
\end{eqnarray}
Up to higher order corrections in $\alpha_s$, $\mu$ dependence of
$T^H(u,r,\mu)$ compensates that of $\varphi_{\pi}(u,\mu)$. It should
be understood that in the above operations the factorization and
renormalization scales have been set identical for simplicity.

Having in hand the hard kernel available, we can calculate the QCD
spectral function to write $F^{QCD}((p+q)^2)$ as a dispersion
integral. For $r=(p+q)^2/m_b^2=s/m_b^2>1$, we have
\begin{eqnarray}
\rho^{QCD}(s)=-\frac{1}{\pi
r}f_{\pi}\int_0^rd\eta~\mathrm{Im}T^H(u,r,\mu)~\varphi_{\pi}(u,\mu)|_{u=\eta/r},
\end{eqnarray}
\begin{eqnarray}
\left.\frac{1}{2\pi}{\mathrm{Im}T^H(u,r,\mu)}\right|_{u=\eta/r}&=&-\delta(1-\eta)+\frac {\alpha_{s}C_{F}}{4\pi}\Bigg\{\delta(1-\eta)\bigg[6-3\mathrm{ln}\frac{m_{b}^{2}}{\mu^{2}}\nonumber\\
&-&\frac{7}{3}\pi^{2}-2\mathrm{Li}_{2}(1-r)+2\mathrm{ln}^{2}(r-1)-2\left(\mathrm{ln}r+\frac{1}{r}-1\right)\mathrm{ln}(r-1)\nonumber\\
&-&2\left(4-3\mathrm{ln}\frac{m_{b}^{2}}{\mu^{2}}\right)\left(1+\frac{d}{d\eta}\right)+2\mathrm{ln}(r-1)\left(1-\mathrm{ln}\frac{m_{b}^{2}}{\mu^{2}}\right)\bigg]\nonumber\\
&+&\left.2\theta(\eta-1)\bigg[\frac{4~\mathrm{ln}(\eta-1)}{\eta-1}\right|_{+}
+\left.\frac{1}{\eta-1}\right|_{+}\Big(\mathrm{ln}\frac{r}{(r-1)^{2}}+\frac{1}{r}\nonumber\\&-2&
+\mathrm{ln}\frac{m_{b}^{2}}{\mu^{2}}\Big)-
\frac{1-r}{\eta r}\left(\mathrm{ln}\frac{\eta}{(\eta-1)^{2}}+1-\mathrm{ln}\frac{m_{b}^{2}}{\mu^{2}}\right)\nonumber\\
&-&\frac{1}{r(r-\eta)}\left(\mathrm{ln}u-2\mathrm{ln}\frac{\eta-1}{r-1}\right)-2\frac{\mathrm{ln}\eta}{\eta-1}-\frac{\eta(r-1)-r}{2r\eta^2}\bigg]\nonumber\\
&+&2\theta(1-\eta)\bigg[\left(\mathrm{ln}\frac{r}{(r-1)^{2}}+\frac{1}{r}-\left.\mathrm{ln}\frac{m_{b}^{2}}{\mu^{2}}\right)\frac{1}{\eta-1}\right|_{+}\nonumber\\
&+&\frac{1}{r(r-\eta)}\left(\mathrm{ln}\frac{r}{(r-1)^{2}}+1-\mathrm{ln}\frac{m_{b}^{2}}{\mu^{2}}\right)-\frac{1-r}{r(r-\eta)}\bigg]
\Bigg\},
\end{eqnarray}
where we take the operation
\begin{eqnarray}
\left.\frac{F(\eta)}{1-\eta}\right|_+=\frac{F(\eta)-F(1)}{1-\eta},
\end{eqnarray}
to avert the redundant divergences possibly occurring as the
integral in (33) is performed over the interval $[0,r]$.

Using (33) and counting the twist-4 contribution covered in (20), we
have the final sum rule for the product $f_Bf_+^{B\to \pi}(0)$
\begin{eqnarray}
f_Bf_+^{B\to\pi}(0)e^{-\frac{m_B^2}{M^2}}&=&-\frac{m_b^2f_{\pi}}{2\pi
m_B^2}\int^{s_{0}^{B}}_{m_{b}^{2}}ds~e^{-\frac{s}{M^2}}\protect\frac{1}{
s}\int_0^{s/m_b^2}d\eta~\mathrm{Im}T\Big(\frac{m_b^2}{s}\eta,\frac{s}{m_b^2},\mu\Big)~\varphi_{\pi}\big(\frac{m_b^2}{s}\eta,\mu\big)\nonumber\\
&+&\frac{f_{\pi}}{m_B^2}\int^{1}_{u_{0}}du
e^{-\frac{m_{b}^{2}}{uM^{2}}}\left
(-\frac{u}{4}\frac{d^{2}\phi_{4\pi}(u)}{du^{2}}\right.\nonumber\\
&+&\left.u\psi_{4\pi}(u)+\int^{u}_{0}dv\psi_{4\pi}(v)-\frac{d}{du}I_{4\pi}(u)\right)\nonumber\\
&\equiv& K(s_0^B, M^2),
\end{eqnarray}
with $M^2$ indicating the Borel parameter with respect to $(p+q)^2$
and $u_{0}=m_{b}^{2}/s_{0}^{B}$.

Converting (36) into the corresponding sum rule for $D\to \pi$
transition by a simple replacement of the parameters, we put an end
to our derivation of the LCSRs for $f^{B\to\pi}_+(0)$ and $f^{D\to
\pi}_+(0)$, to $\mathcal{O}(\alpha_s)$ precision in twist-2
approximation and at tree-level for twist-4 contributions.

We close this Section with a few remarks. Albeit the LCSR
calculations are done at twist-4 level, the results remain valid to
twist-5 accuracy. The reason is simple. The twist-5 DAs as well as
twist-3 ones play no role in the present context due to the Dirac
structures of the related nonlocal operators, of which both
$\bar{d}(x)\gamma_5u(0)$ and
$\bar{d}(x)\sigma_{\mu\nu}\gamma_5u(0)$, as sandwiched between the
vacuum and a pion state, bring about a chirally enhanced twist
expansion. Apart from helping reduce sum rule pollution by
long-distance parameters, the disappearance of twist-3 and -5
components from the light cone expansions guarantees the resulting
LCSRs well convergent. We are going to return to this point in the
following Section.

\section{Choice of the inputs and numerical discussion}

Presently, theoretical estimates of $f^{B\to\pi}_+(0)$ and
$f^{D\to\pi}_+(0)$ with twist-5 accuracy are obtainable in the sum
rules to have been given and the inputs to properly be selected. On
the experimental side, from the measured shapes of the form factors
for $B\to\pi l\tilde{\nu}$, the CKM matrix element $|V_{ub}|$
multiplied by $f^{B\to\pi}_+(0)$ is numerically inferred
as\cite{P.del 052011}:
\begin{eqnarray}
f^{B\to\pi}_+(0)|V_{ub}|=(9.4\pm0.3\pm0.3)\times10^{-4}.
\end{eqnarray}
For the semileptonic processes $D\to\pi l\tilde{\nu}$, a similar
manipulation \cite{D.Besson} gives
\begin{eqnarray}
f^{D\to\pi}_+(0)|V_{cd}|=0.150\pm0.004\pm0.001.
\end{eqnarray}
Then the yielded theoretical predictions could have $|V_{ub}|$ and
$|V_{cd}|$ extracted from these up-to-date data.

Aimed at determining $|V_{ub}|$ and $|V_{cd}|$, we must do our best
to enhance reliability of the LCSR assessments for the form factors
in question. So special care should be taken when making our choice
of the parameters entering the sum rules. The main sources of
uncertainty are, of course, the related DAs, which can merely be
understood at a phenomenological level. Based on the conformal
symmetry of massless QCD, we can parameterize these DAs by expanding
them in terms of matrix elements of conformal operators. The twist-2
DA $\varphi_{\pi}(u)$ is of the following expansion in the
Gegenbauer polynomials:
\begin{eqnarray}
\varphi_{\pi}(u)=6u\bar{u}\Big(1+a_{2}(\mu)C_{2}^{3/2}(u-\bar{u})+a_{4}(\mu)C_{4}^{3/2}(u-\bar{u})+\cdot\cdot\cdot\Big),
\end{eqnarray}
with the even moments $a_{2n}(\mu)$ remaining to be determined. The
Gegenbauer polynomials of higher-degree (large n) are rapidly
oscillating and so one neglects usually their effects on the
numerical integrals included in the sum rules by retaining only the
first few terms of the expansion. Some scenarios have been put
forward to examine the higher-moment effects. We are willing to
mention the prescriptions suggested in \cite{BT} and in
\cite{Huang2}. In \cite{BT} Ball and Talbot (BT) presume that
$a_{2n}$ fall off as powers of $n$, $a_{2n}\propto 1/(n+1)^p$, in
order to build a DA model. In comparison, authors of \cite{Huang2}
consider a modified transverse momentum $\mathbf{K}_{\bot}$
dependent Brodsky-Huang-Lepage (BHL) wavefunction,
\begin{eqnarray}
\Psi_{\pi}(u,\mathbf{K}_{\bot})&=&[1+B_{\pi}C_2^{3/2}(2u-1)+C_{\pi}C_4^{3/2}(2u-1)]\nonumber\\
&\times&\frac{A_{\pi}}{u(1-u)}\mathrm{exp}\left[-\beta_{\pi}^2\left(\frac{\mathbf{K}_{\bot}^2+m_q^2}{u(1-u)}\right)\right],
\end{eqnarray}
which is integrated over $|\mathbf{K}_{\bot}|\leq\mu$ to give a
twist-2 DA. Phenomenological studies with both models are in support
of the rationality of using an expansion truncated after $n=2$. We
stick to such disposal. In one-loop approximation taken as default
for all the renormalized parameters except QCD coupling, $a_2(\mu)$
and $a_4(\mu)$ respect the renormalization group equations
\begin{eqnarray}
a_{2}(\mu_{2})=[L(\mu_{2},\mu_{1})]^{\frac{25C_{F}}{6\beta_{0}}}a_{2}(\mu_{1}),
\end{eqnarray}
\begin{eqnarray}
a_{4}(\mu_{2})=[L(\mu_{2},\mu_{1})]^{\frac{91C_{F}}{15\beta_{0}}}a_{4}(\mu_{1}),
\end{eqnarray}
with
$L(\mu_{2},\mu_{1})=\frac{\alpha_{S}(\mu_{2})}{\alpha_{S}(\mu_{1})}$,
$C_F=4/3$ and $\beta_{0}=11-\frac{2n_{f}}{3}$, $n_f$ being the
number of active quark flavors. To our knowledge, all the existing
estimates for $a_2(\mu)$ are basically consistent with each other
and have the averaged central value of $0.25$ at
$\mu=1~\mathrm{GeV}$. In the light of the current experimental
constraints imposed on LCSR calculations, $a_2(1~\mathrm{GeV})$
appears to prefer varying between $0.16-0.19$
\cite{pballplb,G.Duplancic08, A.Khodjamirian11, Braun}. The
situation is not optimistic about $a_4(\mu)$. The findings differ
among the various studies to a large extent, and even there would be
a difference in sign between numerical estimates. Fortunately, the
sum rule results depend less sensitively on $a_4(\mu)$ than on
$a_2(\mu)$. We would like to use as a consistent input the findings
\cite{A.Khodjamirian11}, $a_2(\mu=1\mathrm{GeV})=0.17\pm0.08$ and
$a_4(\mu=1\mathrm{GeV})=0.06\pm0.1$, from fitting the LCSR
calculation of the pionic electromagnetic form factor to the recent
experimental observation. Concerning the twist-4 DAs, the
three-particle components are specified by only two parameters to
NLO in conformal spin, and are of the following forms:
\begin{eqnarray}
&&\Phi_{4\pi}(\alpha_{i})=120\delta_{\pi}^{2}\varepsilon_{\pi}(\alpha_{1}-\alpha_{2})\alpha_{1}\alpha_{2}\alpha_{3},\\
\nonumber\\
&&\Psi_{4\pi}(\alpha_{i})=30\delta_{\pi}^{2}(\mu)(\alpha_{1}-\alpha_{2})\alpha_{3}^{2}\Big[\frac{1}{3}+2\varepsilon_{\pi}(1-2\alpha_{3})\Big],\\
\nonumber\\
&&\tilde{\Phi}_{4\pi}(\alpha_{i})=-120\delta_{\pi}^{2}\alpha_{1}\alpha_{2}\alpha_{3}\Big[\frac{1}{3}+\varepsilon_{\pi}(1-3\alpha_{3})\Big],\\
\nonumber\\
&&\tilde{\Psi}_{4\pi}(\alpha_{i})=30\delta_{\pi}^{2}\alpha_{3}^{2}(1-\alpha_{3})\Big[\frac{1}{3}+2\varepsilon_{\pi}(1-2\alpha_{3})\Big],
\end{eqnarray}
where the nonperturbative quantities $\delta_{\pi}^{2}$ and
$\varepsilon_{\pi}$ have the scale dependence
\begin{eqnarray}
\delta_{\pi}^{2}(\mu_{2})=[L(\mu_{2},\mu_{1})]^\frac{8C_{F}}{3\beta_{0}}\delta_{\pi}^{2}(\mu_{1}),\nonumber\\
(\delta_{\pi}^{2}\varepsilon_{\pi})(\mu_{2})=[L(\mu_{2},\mu_{1})]^\frac{10}{\beta_{0}}(\delta_{\pi}^{2}\varepsilon_{\pi})(\mu_{1}),
\end{eqnarray}
and the parameter values \cite{P.Ball07} $\delta_{\pi}^{2}=(0.18\pm
0.06)\mathrm{GeV^2}$ and
$\varepsilon_{\pi}=\frac{21}{8}\omega_{4\pi}$($\omega_{4\pi}=0.2\pm
0.1$) normalized at $1~\mathrm{GeV}$, which are to be adopted as
inputs. Resorting to equation of motion the two-particle components,
without introducing any new parameter, can be understood as
\begin{eqnarray}
\phi_{4\pi}(u)&=&\frac{200}{3}\delta_{\pi}^{2}u^{2}\bar{u}^{2}+8\delta_{\pi}^{2}\varepsilon_{\pi}\{u\bar{u}(2+13u\bar{u})\nonumber\\
&+&2u^{3}(10-15u+6u^{2})\mathrm{ln}u+2\bar{u}^{3}(10-15\bar{u}+6\bar{u}^{2})\mathrm{ln}\bar{u}\},\\
\nonumber\\
\psi_{4\pi}(u)&=&\frac{20}{3}\delta_{\pi}^{2}C_{2}^{\frac{1}{2}}(2u-1).
\end{eqnarray}

The $\overline{MS}$ quark masses $m_b$ and $m_c$ comply with the
proverbial LO evolution equations. The bottomonium \cite{J.H.Kuhn}
and charmonium \cite{J.H.Kuhn,R.Boughezal} sum rule results with
four-loop precision, $\bar{m}_{b}(\bar{m}_{b})=4.164\pm
0.025~\mathrm{GeV}$ and $\bar{m}_{c}(\bar{m}_{c})=1.29\pm
0.03~\mathrm{GeV}$, are applicable well to the present discussion.
As far as QCD coupling goes, we use two-loop running down from
$\alpha_s(M_z)=0.1176\pm 0.002$ \cite{PDG}. Additionally, the
factorization scales are assigned, according to the typical
virtuality of the heavy quarks, as $\mu_b=3~\mathrm{GeV}$ and
$\mu_c=1.5~\mathrm{GeV}$ in the respective cases of $B$ and $D$
mesons.

Among the hadronic parameters are the decay constants $f_B$, $f_D$,
and $f_{\pi}$, apart from the heavy meson masses determined
experimentally \cite{PDG} as $m_{B}=5.279 ~\mathrm{GeV}$ and
$m_{D}=1.865~\mathrm{GeV}$. The value of $f_{\pi}$ is measured at
$f_{\pi}=130.4~\mathrm{MeV}$ \cite{PDG}, from the exclusive
processes $\pi\to \mu\widetilde{\nu}_{\mu}$ and $\pi\to
\mu\widetilde{\nu}_{\mu}\gamma$. Recently, an updated measurement of
$f_D$ has already been reported by the CLEO collaboration
\cite{B.I.Eisenstein}, $f_D=205.8\pm8.9~ \mathrm{MeV}$. However, it
is on the basis of combining the experimental data on $f_D$
multiplied by $|V_{cd}|$,
\begin{eqnarray}
f_{D}|V_{cd}|=46.4\pm2.0~ \mathrm{MeV}
\end{eqnarray}
and the assumption $|V_{cd}|=|V_{us}|=0.2255\pm0.0019$, and hence
could only serve as an input in the sum rule calculation of
$f_+^{D\to\pi}(0)$. Instead of a direct estimate of
$f_+^{D\to\pi}(0)$, we consider the sum rule for the product
$f_Df_+^{D\to \pi}(0)$, which in conjunction with the experimental
numbers (38) and (50) allows us to consistently make predictions for
the quantities $f_+^{D\to\pi}(0)$, $|V_{cd}|$ and $f_D$ as well. By
contrast, leptonic $B$ decays are made difficult to detect
experimentally by higher helicity suppression. To have a measurement
analogous to (50), the only opportunity is furnished by $B\to
\tau\nu_{\tau}$ well established lately \cite{tau}. Nevertheless the
results yielded in the SM are less persuasive. The reason is that
these modes turn out to be sensitive to possible extensions of the
SM such as the two-Higgs doublet models and minimal supersymmetric
extensions. We must have recourse to theoretical predictions for
$f_{B}$ to make an assessment of $f_+^{B\to\pi}(0)$. As a consistent
choice, here we make use of the interval $f_{B}=214^{-5}_{+7}$ MeV
\cite{G.Duplancic08} from a sum rule with the $\overline{MS}$ quark
mass.

The remaining parameters are intrinsic to the sum rules, containing
the effective threshold $s_0^B$ $(s_0^D)$ and Borel variables $M^2$.
The former can be set at the neighborhood of the squared mass of the
lowest scalar $B$ meson ($D$ meson). An alternative manner, which
has proven to be more effective, is through use of an auxiliary sum
rule obtained, for example, by taking logarithmic derivative of
$1/M^2$ for (36),
\begin{eqnarray}
m_B^2=-\frac{\partial}{\partial M^{-2}}\mathrm{ln}K(s_0^B, M^2).
\end{eqnarray}
Requiring the measured value of the $B$ meson mass to be reproduced
precisely from the above sum rule, we get the effective interval
$s_{0}^{B}=(34\pm0.5)\mathrm{GeV}^{2}$ in accordance with the sum
rule estimate in heavy quark effective theory \cite{HQCD}.
Similarly, $s_{0}^{D}$ is fixed at $(6.5\pm0.25)\mathrm{GeV}^{2}$.
The Borel intervals could be specified in the standard procedure. We
have $M^2=(18\pm3)~\mathrm{GeV}^{2}$ and
$M^2=(6\pm3)~\mathrm{GeV}^{2}$, corresponding to, respectively, the
sum rules for $B$ and $D$ mesons. As both inherent parameters vary
within their separate ranges allowed, it is demonstrated that the
twist-4 effects are kept at a numerical level less than $4\%$, and
also the continuum contributions are highly suppressed, not
exceeding $20\%$.
\begin{figure}
\vspace{-1.5cm}
\subfigure[]{ \label{fig:mini:subfig:a} 
\begin{minipage}[t]{0.5\textwidth}
\centering
\includegraphics[scale=0.9]{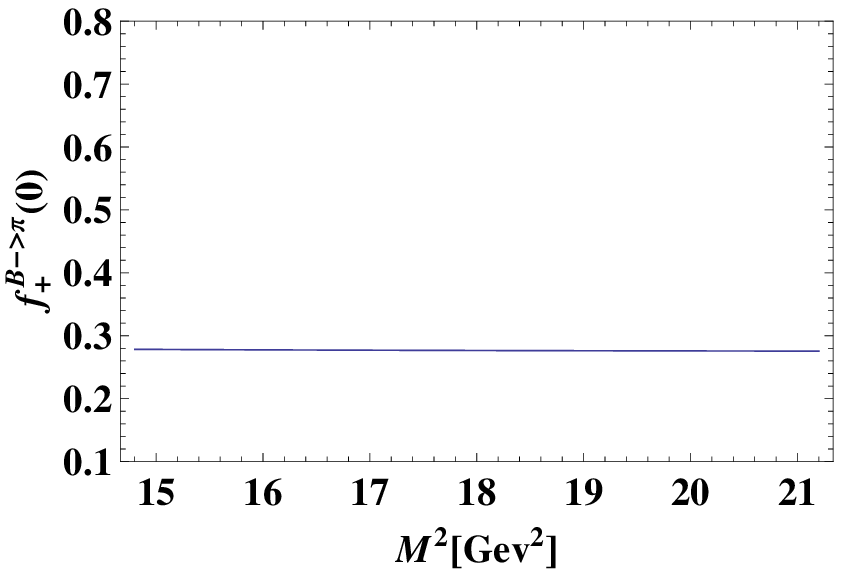}
\end{minipage}}%
\subfigure[]{
\label{fig:mini:subfig:b} 
\begin{minipage}[t]{0.5\textwidth}
\centering
\includegraphics[scale=0.9]{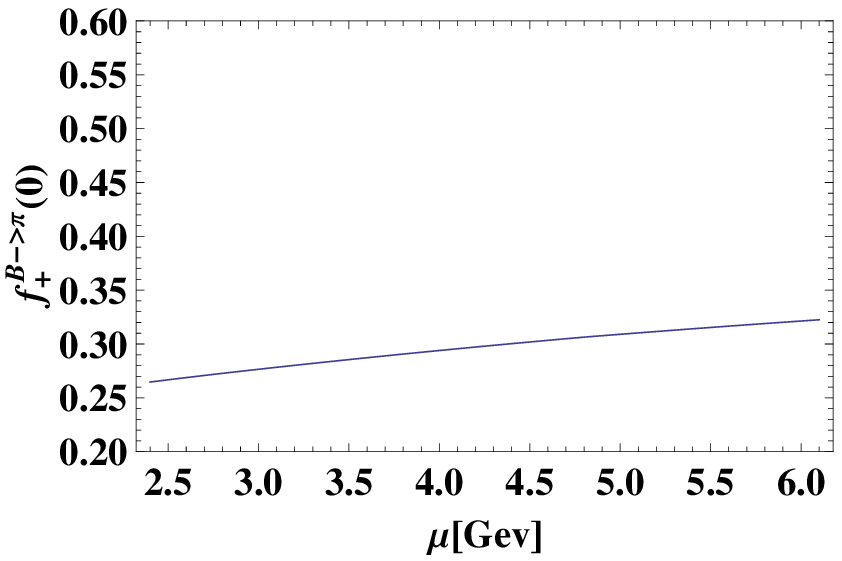}
\end{minipage}}
\begin{center}
\caption{Dependence of the LCSR for $f_{B\pi}^{+}(0)$ on the Borel
parameter $M^2$ (a) and on the factorization scale $\mu$ (b)
\label{fig:(a)} }
\end{center}
\end{figure}

Using the inputs given above, the numerical discussion can be done.
Our sum rule result for $f_{B}f_+^{B\to \pi}(0)$ reads
\begin{eqnarray}
f_{B}f_+^{B\to \pi}(0)=59^{+10}_{-4}~\mathrm{MeV},
\end{eqnarray}
with the uncertainty achieved by adding in quadrature all the errors
caused by variations of the inputs, of which the scale parameter
$\mu$ is set to the interval between $(2.5-6.0)~\mathrm{GeV}$. We
address this result is because it is independent of the value for
$f_{B}$ and therefore of less uncertainty, and moreover is
convenient for a numerical update of the sum rule for $f_+^{B\to
\pi}(0)$ once the theoretical estimate of $f_{B}$ gets improved in
the future. Substituting the parameter value for $f_B$ into (52), we
obtain
\begin{eqnarray}
f_+^{B\to \pi}(0)=0.28^{+0.05}_{-0.02}.
\end{eqnarray}
Illustrating stability of the numerical result, we display the
variations of the sum rule for $f_+^{B\to \pi}(0)$ with the Borel
and the scale parameters, respectively, in Figs.3(a) and 3(b). It is
distinctly observed that the $M^2$ dependence is considerably weak
in the Borel interval required, and there is a moderate $\mu$
dependence. Furthermore, to have an explicit understanding of the
role that every source of uncertainty plays in the uncertainty
evaluation, we collect in Tab.1 the individual uncertainty
contributions estimated by altering each of the inputs within its
specified range. Those not listed therein are tiny and included in
the total uncertainty.
\begin{table}
\centering \caption {The LCSR result for
$f_{+}^{B\rightarrow\pi}(0)$ with the uncertainty estimates due to
the variation of the input.\vspace{0.7cm}}
\begin{tabular}{|c|c|c|c|c|c|c|c|}
\hline
Central value &$ M^{2}$ & $s_{0}^{B}$ &$\mu $ &$m_{b}$&$f_{B}$&$a_{2}^{\pi}$&$a_{4}^{\pi}$\\
\hline
$f_{+}^{B\rightarrow\pi}(0)$  & +0.002 & +0.007&+0.05&+0.008&+0.007&+0.008&+0.01\\
\cline{2-8}
0.277&-0.001& -0.008&-0.01 &-0.008&-0.009&-0.008&-0.01\\
\hline
\end{tabular}
\end{table}
A comparison is drawn among the LCSR predictions for $f_+^{B\to
\pi}(0)$ in Tab.2, there being a result quite close to one another.
We can understand it as follows: (1) No matter which of the two
correlation functions one adopts for a LCSR estimate of that
quantity, the light-cone expansion reveals a good convergence, as
will be addressed. (2) All these calculations employ essentially the
same inputs for the leading twist-2 DA, along with a $f_B$
consistently determined from the sum rules. It is exceptionally hard
to have a LQCD calculation to compare with, since the pionic energy
goes beyond the restriction by the lattice spacing. Nonetheless, it
is claimed \cite{A. Al-Haydari} that $f_+^{B\to \pi}(0)$ is
estimable in an improved LQCD simulation, with the result $f_+^{B\to
\pi}(0)=0.27\pm0.07\pm0.05$.

Now the experimental measurement (37), with the aid of the
theoretical prediction (53), allows for extracting the desired CKM
matrix element $|V_{ub}|$. We have the interval:
\begin{eqnarray}
|V_{ub}|=(3.4^{+0.2}_{-0.6}\pm0.1\pm0.1)\times10^{-3},
\end{eqnarray}
where the first error originates from the uncertainty of $f_+^{B\to
\pi}(0)$ and the others do from the corresponding experimental ones.
Obviously, an analogous result can be extracted in the other LCSR
estimates of $f_+^{B\to \pi}(0)$ in Tab.2. There is also a similar
determination from matching the LCSR calculations and experimental
partial rates for $q^2\leq 12 \mathrm{GeV}^2$
\cite{A.Khodjamirian11}. All these are upheld by the findings in
LQCD simulations for a high $q^2$ and consistent with the CKM fit
upshots \cite{ckmf,utf}.
\begin{table}
\centering \caption {Comparison of theoretical predictions for the
form factors $f_{+}^{B\rightarrow\pi}(0)$ and
$f_{+}^{D\rightarrow\pi}(0)$.\vspace{0.7cm}}
\begin{tabular}{|c|c|c|}
\hline
~~~~~~~~~Approach~~~~[Ref.]~~~~~~~~~&~~~~~~~~~$ f_{+}^{B\rightarrow\pi}(0)$~~~~~~~~~ &~~~~~~~~~$ f_{+}^{D\rightarrow\pi}(0)$ ~~~~~~~~~\\
\hline
LCSR~~~~~~~~~\cite{A.Khodjamirian11}&$0.281\pm0.05$&\\
~~~~~~~~~~~~~~~~~\cite{pball05}&$0.258\pm 0.331$& \\
~~~~~~~~~~~~~~~~~\cite{P.Ball06}&&$0.63\pm 0.11$ \\
~~~~~~~~~~~~~~~~~\cite{G.Duplancic08}&$0.26^{+0.04}_{-0.03}$&\\
~~~~~~~~~~~~~~~~~\cite{A.Khodjamirian09}&&$0.67^{+0.10} _{-0.07}$\\
~~~~~~~~~~~This work&$0.28^{+0.05}_{-0.02}$&$0.62\pm0.03$\\
 \hline
 Lattice QCD \cite{A. Abada}&&$0.57\pm0.06\pm0.02$\\
~~~~~~~~~~~~~~~~~~\cite{C.Aubin}&&$0.64\pm0.03\pm0.06$\\
~~~~~~~~~~~~~~~~~~\cite{A. Al-Haydari}&&$0.74\pm0.06\pm0.04$\\
~~~~~~~~~~~~~~~~~~\cite{Hee Na}&&$0.666\pm0.029$\\
~~~~~~~~~~~~~~~~~~\cite{S.Di}&&$0.65\pm0.06\pm0.06$\\
\hline
\end{tabular}
\end{table}

Corresponding to (52), the product $f_{D}f_+^{D\to \pi}(0)$ has the
numerical value
\begin{eqnarray}
f_{D}f_+^{D\to \pi}=117^{+8}_{-7}~\mathrm{MeV}.
\end{eqnarray}
Tab.3 provides a summary of the major uncertainty contributions to
the sum rule. As exhibited in Figs.4(a) and 4(b), the stability of
the sum rule holds as well as in the B meson situation, as $M^2$
changes in the interval specified and $\mu$ ranges from 1 to 3
$~\mathrm{GeV}$. Intriguingly, using the same inputs as ours for
most of the parameters this quantity is explored in the LCSR
approach \cite{A.Khodjamirian09} and the yielded result
$f_{D}f_{D\pi}^{+}(0)=137^{+19}_{-14} ~\mathrm{MeV}$ is compatible
with our prediction within the errors, but showing a larger central
value. We remark on this difference. The twist expansion in
$x^2\approx0$ is the basic thought of the LCSR approach. For heavy
to light transition, such an expansion must match the one in the
inverse of heavy quark mass $m_Q$. One shows, indeed, that in the
heavy quark expansion the end point behaviors of the higher-twist
DAs entering a traditional LCSR might modify, but does not violate
the twist hierarchy. For instance, the twist-3 term, which is
formally $1/m_Q$ suppressed versus the twist-2 part, behaves the
same as the latter in the heavy quark limit. However, an explicit
calculation with a finite $m_Q$ demonstrates that whereas the twist
expansion works better for $B$ decays, there is a considerable
numerical violation of the hierarchy relation in the $D$ meson
cases, where the twist-3 components contribute to the sum rules much
more than the twist-2 ones due to the chiral enhancement factor
$\mu_{\pi}>1$. The fact that the sum rule for $f_+^{D\to \pi}(0)$ is
poorly convergent implies that the twist-5 effect is not negligible
and should be considered, even if we work in twist-4 approximation.
Currently nothing is known, however, about the twist-5 DAs except
that they provide the sum rule with a term formally $1/m_c^2$
suppressed with respective to the twist-3 one. To have a sketchy
understanding of their influence on the LCSR calculation, authors of
\cite{A.Khodjamirian09} suppose that the ratio of the twist-5 to -3
parts is identical to the one of the twist-4 and -2 terms, while in
\cite{P.Ball06} the twist-4 term is multiplied by a factor of 3.
Anyway, it is still obscure that how much on earth do the twist-5
components, in particular those with the chiral enhancement factor,
contribute to the sum rule for $f_+^{D\to \pi}(0)$. We leave it as
an open question until a reliable twist-5 model wavefunction is
presented. Given that the present scenario ensures, to twist-5
precision, the light-cone expansion to converge well whether for $B$
or $D$ decays, this issue gets, at any rate, settled provisionally.
\begin{figure}
\subfigure[]{ \label{fig:mini:subfig:a} 
\begin{minipage}[t]{0.5\textwidth}
\centering
\includegraphics[scale=0.9]{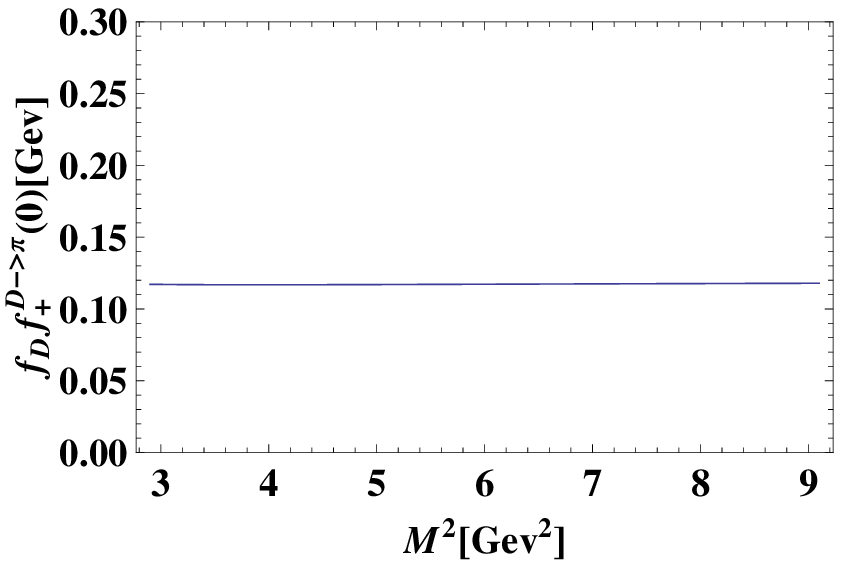}
\end{minipage}}%
\subfigure[]{
\label{fig:mini:subfig:b} 
\begin{minipage}[t]{0.5\textwidth}
\centering
\includegraphics[scale=0.9]{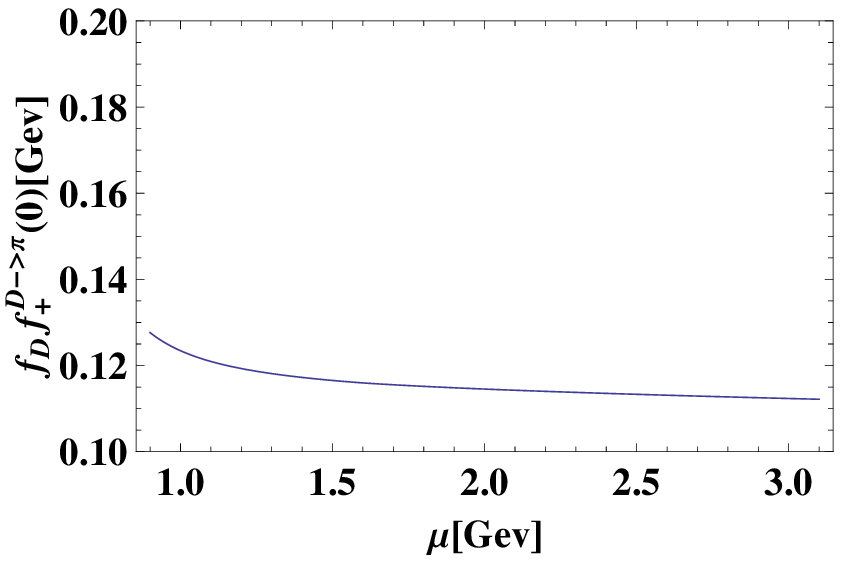}
\end{minipage}}
\begin{center}
\caption{Dependence of the LCSR for $f_{D}f_{D\pi}^{+}(0)$ on the
Borel parameter $M^2$ (a) and on the factorization scale $\mu$ (b).}
\end{center}
\end{figure}

\begin{table}
\centering \caption {The LCSR result for
$f_{D}f_{+}^{D\rightarrow\pi}(0)$ with the uncertainty estimates due
to the variation of the input.\vspace{0.7cm}}
\begin{tabular}{|c|c|c|c|c|c|c|c|c|}
\hline
Central value &$ M^{2}$ & $s_{0}^{D}$ &$\mu $ &$m_{c}$&$a_{2}^{\pi}$&$a_{4}^{\pi}$&$\omega_{4}^{\pi}$&$\delta_{\pi}^{2}$\\
\hline
$f_{D}f_{+}^{D\rightarrow\pi}(0)$  & +0.0007 & +0.0015&+0.0062&+0.0001&+0.005&+0.0017&+0.0003&+0.0014 \\
\cline{2-9}
0.117&-0.0001& -0.0016&-0.0049 &-0.0003&-0.005&-0.0016&-0.0002&-0.0013\\
\hline
\end{tabular}
\end{table}

Let us go back to our numerical calculation. Combining the sum rule
prediction (55) with the product of the two experimental numbers
(38) and (50), we could yield the square of $|V_{cd}|$ and further
the magnitude of $V_{cd}$:
\begin{eqnarray}
|V_{cd}|=0.244\pm0.005\pm0.003\pm0.008,
\end{eqnarray}
where the first and second errors are of an experimental origin and
the third is due to the theoretical uncertainty. This result
deviates by about $2\%$ from the Wolfenstein approximation
$|V_{cd}|= |V_{us}|=0.2255\pm0.0024$ \cite{PDG}, and is in good
keeping with $|V_{cd}|=0.234\pm0.007\pm0.002\pm0.025$ extracted from
(38) by using the LQCD estimate $f_{D\pi}^{+}(0)=0.64\pm0.03\pm0.06$
\cite{C.Aubin}. Certainly we have a slightly larger central value
than achieved in \cite{A.Khodjamirian09}, where the same data are
combined with the LCSR result $f_{D}f_+^{D\to
\pi}(0)=137^{+19}_{-14}$~MeV.

To proceed, substitution of (56) in (38) gets
\begin{eqnarray}
f_{+}^{D\rightarrow\pi}(0)=0.62\pm0.03,
\end{eqnarray}
where all the theoretical and experimental errors are in quadrature
covered in the total uncertainty. There are abundant researches on
$f_{+}^{D\rightarrow\pi}(0)$, from which we pick just out several
typical LCSR and lattice predictions and arrange them, along with
the present estimate, into the tabulation in Tab.2. At first sight,
there exists a good accordance among all the LCSR results listed.
Yet this should not be taken too seriously, for more or less
parameters, on which the sum rules have relatively sensitive
dependence, are chosen to have different inputs in these
calculations. For instance, the obviously different parameter values
are employed for both $f_D$ and $a_2$ in \cite{P.Ball06} and
\cite{A.Khodjamirian09}. The LQCD evaluations turn out to have a
different extent of deviation from each other in the central values,
but without any conflict within the errors. A comprehensive survey
shows that $f_{+}^{D\rightarrow\pi}(0)$ prefers taking a value
larger than $0.6$.

Lastly, we would like to present, as a by-product, our assessment
for the decay constant $f_D$. From (50) and (56) follows that
\begin{eqnarray}
f_{D}=190^{+12}_{-11}~\mathrm{MeV},
\end{eqnarray}
with the same error disposal as in the $f_{+}^{D\rightarrow\pi}(0)$
case. It falls into a somewhat wide interval formed by the existing
findings of $f_{D}$, which can be illuminated by the following
examples. The CLEO measures $f_D=205.8\pm8.9~\mathrm{MeV}$
\cite{B.I.Eisenstein} on the assumption $|V_{cd}|=
|V_{us}|=0.2255\pm0.0019$. LQCD simulation predicts the three-flavor
results $f_D=218.9\pm11.3~\mathrm{MeV}$ \cite{Fermilab} and
$f_D=213\pm4~\mathrm{MeV}$ \cite{HPQCD}, and two-flavor one
$f_D=197\pm4~\mathrm{MeV}$ \cite{B.Blossier}. Compared with all
these determinations, QCD sum rules provide, besides the two-loop
result $203\pm20~\mathrm{MeV}$ \cite{S.Narison}, the three-loop ones
$f_D=195\pm20~\mathrm{MeV}$ \cite{A.penin} and
$f_D=177\pm21~\mathrm{MeV}$ \cite{J.Bordes}. Hence one should step
up efforts to improve calculations and promote understanding of that
quantity. The present estimate, however, could be accommodated by
$f_D\approx200~\mathrm{MeV}$, a result gradually becoming accepted
on the basis of a multitude of phenomenological investigations, and
in particular accords well with those from the two-flavor LQCD
\cite{B.Blossier} as well as three-loop QCD sum rules
\cite{A.penin}. Meanwhile, these consistencies further expand
support for the validity of our findings in (56) and (57).

In the above discussion, a cross check has been made automatically
between our LCSR predictions for $f_{+}^{B\rightarrow\pi}(0)$ and
$f_{+}^{D\rightarrow\pi}(0)$. In contrast, it is out of the question
for a traditional LCSR calculation, since twist-2 and -3
contributions, as emphasized, dominate respectively in the two sum
rules, which consequently show exceedingly different sensitivities
to both of them. In addition, from the observation that the twist-2
part predominates entirely over the twist-4 one in the present LCSR
framework, we can benefit a lot in attempting to acquire a
constraint on $a_2$ and $a_4$ from the data on $B$ and $D$ decays.
No doubt, this would enhance significantly our confidence in the
LCSR applications to heavy-to-light transitions.

\section{Summary}

We have addressed in some detail a QCD assessment for $B,~D\to\pi$
transitions at the zero momentum transfer, in an improved LCSR
approach, and presented our determinations of the form factors
$f_{+}^{B\rightarrow\pi}(0)$ and $f_{+}^{D\rightarrow\pi}(0)$ as
well as the CKM matrix elements $|V_{ub}|$ and $|V_{cd}|$. We have
also yielded a numerical estimate of the decay constant $f_{D}$.

To $\mathcal{O}(\alpha_s)$ accuracy for twist-2 contributions and
with the $\overline{MS}$ masses for the heavy quarks, the LCSR
calculation on $f_Bf_{+}^{B\rightarrow\pi}(0)$ and
$f_Df_{+}^{D\rightarrow\pi}(0)$ is carried out and the resulting sum
rules bear the two remarkable characteristics: (1) They receive no
contribution from not only the twist-3 but also the unknown twist-5
components, which are regarded usually as a serious source of
uncertainty in the conventional LCSR applications, among others, to
$D$ decays, and therefore are available to twist-5 accuracy. (2) The
twist-2 parts play a fully dominant role over the twist-4 ones so
that the twist hierarchy required for convergence of the light cone
expansions is preserved well and the higher-twist effects are kept
under good control. The numerical analysis is performed with the
updated inputs and experimental data; the validity and the
self-consistency of the sum rule results are checked up and verified
by a numerical comparison with some of typical theoretical
predictions. Our findings are such as below:
\begin{eqnarray}
&&f_{+}^{B\rightarrow\pi}(0)=0.28^{+0.05}_{-0.02},~~~~~~|V_{ub}|=(3.4^{+0.2}_{-0.6}\pm0.1\pm0.1)\times10^{-3},\nonumber\\
\nonumber\\
&&f_{+}^{D\rightarrow\pi}(0)=0.62\pm0.03,~~~|V_{cd}|=0.244\pm0.005\pm0.003\pm0.008,\nonumber\\
\nonumber\\
&&f_{D}=190^{+12}_{-11}~\mathrm{MeV}.\nonumber
\end{eqnarray}

The present results can be improved once the related inputs or
experimental data become updated. Albeit unlikely to give help in
understanding the existing discrepancy between inclusive and
exclusive $|V_{ub}|$ determinations, an improvement on the
$|V_{ub}|$ determination is expected especially. However, it demands
evidently a more decided knowledge of $f_B$, apart from a
significant advance in experiment and in theoretical or
phenomenological research on the pionic twist-2 DA. A continued and
intensive study of $f_B$ helps also in the confirmation whether or
not non-SM physics shows an explicitly observable effect in
$\tau$-leptonic and corresponding semileptonic $B$ decays, which are
expected to be detectable to a high precision in the running LHC or
foreseeable super $B$ factor. On the other hand, although so far our
discussion on the form factors has been restricted to the largest
recoil point $q^2=0$, $q^2$ dependence of them is understandable
within the kinematical regions allowed by their individual
light-cone expansion calculations. Then it is possible to
extrapolate the results to the large $q^2$ regions in various ways
available so as to have an all-around understanding of their
behaviors. Too it is interesting to generalize the present
discussion to the decays into $K$ meson. We put off these studies to
a future issue.

\section*{Acknowlegements} N.~Zhu would like to thank
Dr.~Y.~-M.~Wang for helpful discussion in the numerical calculation.
This work is in part supported by the National Science Foundation of
China under Grant Nos.10675098 and 11175151.

\end {document}